\documentclass{ws-procs10x7}
\usepackage{balance}
\usepackage{epsfig}

\newcolumntype{d}[1]{D{.}{.}{#1}}

\def\Journal#1#2#3#4{{\it #1} {\bf #2}, #3 (#4)}

\newcommand{\PO}{\rm l \! P }

\newcommand{\xpom}{x_{\PO} }

\makeindex
\begin{document}

\title{QCD FITS TO DIFFRACTION\,$^*$}

\author{S. SAPETA$^{\dagger}$} 
\address{M. Smoluchowski Institute of Physics, Jagellonian University, \\
         Reymonta 4, 30-059 Krak\'ow, Poland}


\twocolumn[\maketitle\abstract{
The most recent H1 and ZEUS data for diffractive structure functions are analyzed under three different theoretical approaches. This includes the Pomeron Structure Function (PSF) framework, Bartels-Ellis-Kowalski-W\"usthoff (BEKW) color dipole approach and the Golec-Biernat-W\"usthoff (GBW) saturation model. The models are shown to successfully fit the set of combined  data. Conceptual differences between the models are discussed and, as an example, the prediction for the longitudinal diffractive structure function is presented.
}
\keywords{diffraction; deep inelastic scattering; 
          pomeron; color dipole; saturation.}
]

\section{Introduction}

\footnotetext{$^{*\,}$Presented at the XXXIII International Conference on High Energy Physics, Moscow, Russia, July 26 -- August 2, 2006.}
\footnotetext{$^{\dagger\,}$Present address: Department of Physics, CERN, Theory Division, CH-1211 Geneva 23, Switzerland, e--mail: Sebastian.Sapeta@cern.ch}

The process of inelastic electron--proton diffraction, $ep \to eXY$, have been widely studied over last years at the HERA collider.
Quite recently both H1 and ZEUS collaborations published particularly precise data
[\refcite{h1rapdata}--\refcite{zeustagdata}]. 
In this short letter we present the results of the analysis of this data in the frameworks of three different theoretical models. 
These are: the Pomeron Structure Function (PSF) model formulated in the framework of Regge phenomenology, the Bartels-Ellis-Kowalski-W\"usthoff (BEKW) two gluon exchange dipole model and the saturation model of Golec-Biernat and W\"usthoff (GBW).
By fitting each of these models to the combined set of all measurements we test also the compatibility of the data obtained with various experimental methods.

When we plot the experimental data  for the diffractive structure function
\mbox{[\refcite{h1rapdata}--\refcite{zeustagdata}]}
we observe that it decreases in the range of low $\xpom$ values, reaches a minimum at certain point, and starts to grow up as $\xpom$ goes to 1. Thus, $F_2^{D(3)} (Q^2,\beta, \xpom)$ is expressed as a sum of two components
\begin{eqnarray}
F_2^{D(3)}
& = & 
F_2^{D(3),\; {\rm low-}\xpom} + F_2^{D(3),\; {\rm large-}\xpom}. \quad
\label{reggeform}
\end{eqnarray}
In our analysis we concentrate on the low-$\xpom$ part of the diffractive structure function and test various theoretical frameworks in which this quantity may be calculated. The large-$\xpom$ part is modeled as a Reggeon exchange with the structure function of the pion [\refcite{owens}] and only its normalization for each data set is determined from the fit.

\section{Data sets and fitting method}

\begin{table*}
\tbl{Results, in terms of $\chi^2$, of the fits to the three data sets for the PSF, BEKW and GBW models. \label{tab1}}
{\begin{tabular}{l@{\quad }c@{\qquad }c@{\qquad }c}
\toprule
$\chi^2$/(nb data points) & H1RAP      & ZEUSMX           & All data sets \\ 
\colrule 
PSF                 & 250.3/240 = 1.04 & 100.9/102 = 0.98 & 377.6/444 = 0.85 \\ 
BEKW                & 286.8/247 = 1.16 & 193.5/142 = 1.36 & 493.1/493 = 1.00 \\ 
GBW                 & 272.9/247 = 1.10 & 268.0/142 = 1.88 & 564.5/493 = 1.15 \\ 
\botrule
\label{tab:chi2}
\end{tabular}}
\end{table*}

We use the data for the reduced cross section $\sigma_r^{D(3)} (Q^2,\beta, \xpom)$ or the diffractive structure function $F_2^{D(3)} (Q^2,\beta, \xpom)$  published by H1 [\refcite{h1rapdata}, \refcite{h1tagdata}] and ZEUS [\refcite{zeusmxdata}, \refcite{zeustagdata}]  collaborations, respectively. These data sets are, however, obtained with different methods and therefore for different upper limits on the proton dissociation system  mass $M_Y$. Hence, they need to be normalized to the same experimental situation, $M_Y < 1.6\ {\rm GeV}$, in accordance with~[\refcite{h1rapdata}]. Taking all the above into account we have the following data sets which are used in the fits

\begin{itemlist}
\item $\sigma_r^{D(3)}$ measured by the H1 using the rapidity gap method~[\refcite{h1rapdata}] with $M_Y < 1.6\ {\rm GeV}$ called H1RAP, default data set, not corrected further,

\item $F_2^{D(3)}$ measured by the ZEUS using the $M_X$ method~[\refcite{zeusmxdata}] with $M_Y < 2.3\ {\rm GeV}$ called ZEUSMX, multiplied  by the factor 0.85, 

\item $\sigma_r^{D(3)}$ measured by the H1 with the proton detected in roman pot detectors~[\refcite{h1tagdata}] called H1TAG, multiplied by the factor 1.23,

\item $F_2^{D(3)}$ measured by the ZEUS with the proton detected in roman pot detectors [\refcite{zeustagdata}] called ZEUSTAG, multiplied by the factor~1.23.
\end{itemlist}

We perform three fits for each model. The fit to H1RAP data alone, the fit to ZUESMX data alone and the combined fit to all four data sets listed above. In the first two cases only statistical and uncorrelated systematic uncertainties added in quadrature were used in the calculation of $\chi^2$ whereas the total errors were taken in the third case.
The cut \mbox{$Q^2 > 4.5\ {\rm GeV}^2$} was imposed for all fits.

\section{Pomeron Structure~Function (PSF) model}

In the framework of Regge phenomenology the low-$\xpom$ component of the diffractive structure function is attributed to the Pomeron exchange. The $\xpom$ is factorized from $\beta$ and $Q^2$ dependence and we have 
\begin{eqnarray}
\label{eq:f2psf}
F_2^{D(3),\; {\rm low-}\xpom} = 
f_{\PO / p} (x_{\PO}) F_2^{D(\PO)} (Q^2,\beta),
\end{eqnarray}
where the first factor is called the Pomeron flux. It has the standard form [\refcite{Royon:2006by}] with the Pomeron intercept $\alpha_{\PO}(0)$, which governs the $\xpom$ dependence and is let as a free parameter in our fit.  The second factor $F_2^{D(\PO)}(Q^2, \beta)$ is interpreted as the Pomeron structure function defined in analogy to the proton structure function. In the same manner diffractive parton distribution functions (DPDFs) are defined for the Pomeron. The evolution of these distributions with $Q^2$ is governed by the DGLAP [\refcite{dglap}] equations.

In our analysis we take the following form of singlet $zS$ and gluon $zG$ distributions at initial scale $Q^2_0$ 
\begin{eqnarray}
z{\it {S}}(z,Q_0^2) 
&=&
A_S z^{B_S}(1-z)^{C_S}                   \nonumber \\
& &
\cdot\, (1+D_S z + E_S \sqrt{z})  
\cdot e^{\frac{0.01}{z-1}},              \nonumber \\
z{\it {G}}(z,Q_0^2) &=& 
A_G (1-z)^{C_G}   
\cdot e^{\frac{0.01}{z-1}}.
\label{gluona}
\end{eqnarray}
They are similar to that used in fit A of H1~[\refcite{h1rapdata}] however we increased the number of parameters adding $D_S$ and $E_S$. This allowed us to make the fit independent on the choice of $Q_0^2$, which is fixed at $Q_0^2 = 3.0\ {\rm GeV}^2$. 
Hence, we have eight parameters introduced by the function defined in Eq.~(\ref{eq:f2psf}).
The charm contribution is not fitted but calculated according to [\refcite{ghrgrs}].
Since in this approach the higher twist effects are not present we need to restrict the kinematic range by introducing two additional cuts $M_X = 2.0\ {\rm GeV}$ and $\beta < 0.8$. 
 
The fit results in terms of $\chi^2$ are presented in the Table \ref{tab:chi2}. Very good fit quality is obtained in all three cases. 
The gluon density turns out to dominate in the Pomeron for all fits however its shape varies between fits with different data sets.
We obtain $\alpha_{\PO}(0) = 1.118 \pm 0.005$ in the combined fit which agrees with the result found by H1 (fit A) [\refcite{h1rapdata}].
The full set of parameters as well as the singlet and gluon distributions can be found in [\refcite{Royon:2006by}].

\section{Bartels-Ellis-Kowalski-W\"usthoff~(BEKW)~model}

In this model proposed in [\refcite{bartels}] the photon--proton interaction is realized in two stages. In the first stage the virtual photon splits into $q\bar q$ or $q \bar q g$ Fock states known as color dipoles. Then the dipole interacts with the proton via two gluon exchange. Thus, the simplest model of Pomeron as the gluon pair is assumed.
The low-$\xpom$ component of the diffractive structure function consists of three terms
\begin{eqnarray}
\label{eq:f2bekw}
F_2^{D(3),\; {\rm low-}\xpom} = 
F_T^{q\bar q} + F_L^{q\bar q} + F_T^{q\bar q g}
\end{eqnarray}
where T (L) refers to transverse (longitudinal) polarization of the incoming photon.
From the properties of the wave function the $\beta$ dependence of each term was deduced. The $q\bar q_T$ term dominates  for intermediate $\beta$ values whereas $q\bar q_L$ is the most important at large~$\beta$. Moreover, the longitudinal component behaves like a higher twist part so neither $\beta$ nor $M_X$ cuts are needed. For the low $\beta$ values the $q \bar q g_T$ term is the largest. The $\xpom$ dependence is impossible to guess from perturbative QCD therefore it is assumed in the form 
$(x_0/\xpom)^{n_{2(4)} (Q^2)}$ where 
$n_{2(4)}(Q^2) = n^0_{2(4)} + n^1_{2(4)} \ln[Q^2/Q^2_0 +1]$ with $n^0_{2(4)}$, $n^1_{2(4)}$ to be determined from the fit. Altogether the function given in Eq.~(\ref{eq:f2bekw}) has eight parameters.

In the second row of Table \ref{tab:chi2} we present the $\chi^2$ values resulting from the fit of the BEKW model to the three data sets. Good fit quality is observed in the case of H1RAP and the combined fit. The fit to the ZEUSMX data set gives slightly worse $\chi^2$. For all details concerning the parameter values and plots illustrating $\beta$ dependence of $F_2^{D(3),\; {\rm low-}\xpom}$ in the BEKW model the reader is referred to~[\refcite{Royon:2006by}].

\section{Golec-Biernat-W\"usthoff (GBW) saturation model}

The saturation model proposed in [\refcite{gbw1}, \refcite{gbw2}] was formulated in the color dipole picture.  Hence, the general structure of $F_2^{D(3),\; {\rm low-}\xpom}$ is identical with the BEKW model and therefore given by Eq.~(\ref{eq:f2bekw}). However in this model the cross section for the interaction of the color dipole with  the proton is  assumed in the form
\begin{equation}
\hat\sigma (\xpom,r)\,=\,\sigma_0\,\left\{
1\,-\,\exp\left(-r^2\, Q^2_s(\xpom)/4  \right) \right\},
\end{equation}
where $Q^2_s(\xpom) = \left(x_0/\xpom\right)^{\lambda}$ is called the saturation scale. So, the cross section saturates either for $r\to \infty$ and $\xpom$ fixed or $\xpom \to 0$ and $r$ fixed. We used the version of the GBW model with three massless quarks. The model has only four parameters to be determined from the fit (see [\refcite{Royon:2006by}]). No $\beta$ or $M_X$ cuts are imposed for the same reason as discussed in the previous section. 

We present the results of $\chi^2$ from the GBW fits in the third row of Table~\ref{tab:chi2}. Good quality is obtained in the case of H1RAP and the combined fit. Significantly worse result is found when the GBW model is fitted to  the ZEUSMX data set. The results for parameters, especially $x_0$, vary significantly  depending on the data set used in the fit. Again, we refer the read to [\refcite{Royon:2006by}] for further details.

\section{Conclusions}
We have presented the results of the fits of three different theoretical models to the most recent data on the diffractive structure functions from H1 and ZEUS collaborations. We have shown that the combined fit to the four data sets with total errors is successful, in terms of $\chi^2$, for all models. 
Especially good description of the diffractive data is found in the framework of the PSF and BEKW models.
In addition, the PSF approach works good also when fitted to the ZEUSMX data set, which is not the case for BEKW and GBW models. 

\begin{figure}[t]
\epsfig{file=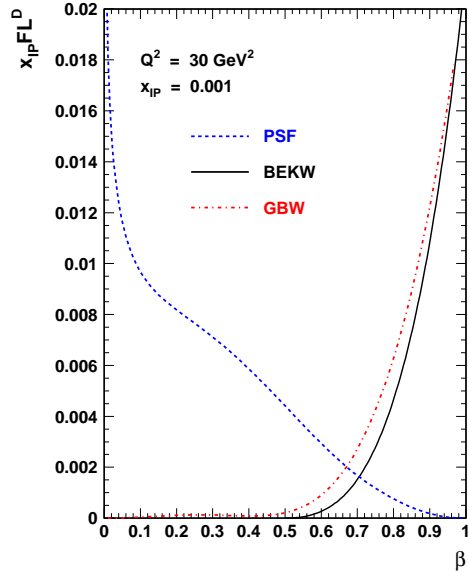,width=7.0cm}
\caption{Predictions for the longitudinal part of the diffractive structure function as a function of $\beta$.}
\label{fig10}
\end{figure}

Let us stress that the three frameworks presented here are based on significantly different theoretical concepts. This is reflected \emph{e.g.} in the twist structure of the models. As we see in Fig.~\ref{fig10}, the predictions for the longitudinal part of the diffractive structure function $F_L^D$ are very different between PSF and BEKW/GBW approaches. 
Since in the PSF approach only the twist--two component is present we expect it to give correct $F_L$ in the low $\beta$ region. On the other hand, because $F_L$ in the BEKW/GBW models has only twist--four part this models are reliable for large~$\beta$.

Finally, let us mention that in [\refcite{Royon:2006by}] the reader may find  
also the results for the Bia\l as-Peschanski (BP) model [\refcite{bpmodel}] based on the BFKL Pomeron approach. The combined fit of this model in the kinematic range as in the case of the PSF model and additional cut $Q^2 < 120\ {\rm GeV}^2$ leads to $\chi^2/{\rm (nb\ data\ points)} = 1.26$.

For more details concerning the analysis described here we refer to the original paper~[\refcite{Royon:2006by}].

\section*{Acknowledgments}
This work was done in collaboration with C.~Royon, L.~Schoeffel, R.~Peschanski and E.~Sauvan.
S.S. acknowledges MEN grants: No. 1 P03B 028 28 (2005-08) and N202 048 31/2647 (2006-08) and the French-Polish agreement Polonium. 



\begin{thebibliography}{9}

\bibitem{h1rapdata}
  A.~Aktas {\it et al.}  (H1 Collaboration),
  hep-ex/0606004.  
\bibitem{h1tagdata}
  A.~Aktas {\it et al.}  (H1 Collaboration),
  hep-ex/0606003 .  
\bibitem{zeusmxdata}
  S.~Chekanov {\it et al.}  (ZEUS Collaboration),
  \Journal{Nucl. Phys.}{B713}{3}{2005}.
\bibitem{zeustagdata}
  S.~Chekanov {\it et al.}  (ZEUS Collaboration),
  \Journal{Eur. Phys. J.}{C38}{43}{2004}.
\bibitem{owens} 
  J.~F.~Owens
  \Journal{Phys. Rev.}{D30}{943}{1984}.
\bibitem{Royon:2006by}
  C.~Royon, L.~Schoeffel, S.~Sapeta, R.~Peschanski and E.~Sauvan, \\
  hep-ph/0609291.
\bibitem{dglap} 
  G.~Altarelli and G.~Parisi,
  \Journal{Nucl. Phys.}{B126}{298}{1977};
  V.~N.~Gribov and L.~N.~Lipatov, 
  \Journal{Sov. Journ. Nucl. Phys.}{15}{438}{1972};
  \Journal{Sov. Journ. Nucl. Phys.}{15}{675}{1972};
  Yu.~L.~Dokshitzer
  \Journal{Sov. Phys. JETP.}{46}{641}{1977}.
\bibitem{ghrgrs} 
  M.~Gl\"uck, E.~Hoffmann, E.~Reya, 
  \Journal{Z.~Phys.}{C13}{119}{1982};
  M.~Gl\"uck, E.~Reya, M.~Stratmann, 
  \Journal{Nucl.~Phys.}{B422}{37}{1994}.
\bibitem{bartels} 
  J.Bartels, J.Ellis, H.Kowalski, M.Wusthoff,  
  \Journal{Eur.Phys.J.}{C7}{443}{1999}.
\bibitem{gbw1}
  K.~Golec-Biernat and M.~Wusthoff,
  \Journal{Phys. Rev.}{D59}{014017}{1999}.
\bibitem{gbw2} 
  K.~Golec-Biernat and M.~Wusthoff,
  \Journal{Phys. Rev.}{D60}{114023}{1999}.
\bibitem{bpmodel} 
  A.~Bialas, R.~Peschanski, C.~Royon, 
  \Journal{Phys. Rev.}{D57}{6899}{1998}; 
  S.~Munier, R.~Peschanski, C.~Royon, 
  \Journal{Nucl. Phys.}{B534}{297}{1998}.

\end{thebibliography}
\end{document}